\documentclass[%
reprint,
superscriptaddress,
amsmath,amssymb,
aps,
prl,
]{revtex4-2}
\usepackage{color}
\usepackage{graphicx}
\usepackage{dcolumn}
\usepackage{bm}
\usepackage{textcomp}

\begin {document}

\title{Tunable Anomalous Diffusion in Subrecoil-Laser-Cooled Atoms}

\author{Soma Shiraki}
\thanks{These authors contributed equally to this work.}
\affiliation{%
  Department of Physics and Astronomy, Tokyo University of Science, Noda, Chiba 278-8510, Japan
}%

\author{Eli Barkai}
\thanks{These authors contributed equally to this work.}
\affiliation{%
  Department of Physics, Bar-Ilan University, Ramat-Gan, Israel
}%


\author{Takuma Akimoto}
\email{takuma@rs.tus.ac.jp}
\affiliation{%
  Department of Physics and Astronomy, Tokyo University of Science, Noda, Chiba 278-8510, Japan
}%



\date{\today}

\begin{abstract}
Laser cooling of atomic motion enables advances in quantum information and precision metrology. 
However, the spatial spreading of subrecoil-laser-cooled atoms---crucial for understanding cooling mechanisms and atomic confinement---remains largely unexplored. Here, we analyze anomalous diffusion in subrecoil-laser-cooled atoms, where a velocity-dependent fluorescence rate  $R(v) \propto |v|^{\alpha}$ governs transport properties. By tuning  $\alpha$, we uncover transitions between normal, subdiffusive, and superdiffusive regimes. Notably, at  $\alpha = 3/2$, diffusion is minimized, leading to optimal atomic confinement.  
We further identify a conceptual link between subrecoil laser cooling and the Pomeau-Manneville map from nonlinear dynamics, revealing that anomalous diffusion can generically exhibit a nontrivial minimum in spatial spreading---even across seemingly unrelated physical systems.
\end{abstract}

\maketitle




Laser cooling techniques have revolutionized atomic motion control, driving advances in both fundamental physics and applied technologies, including quantum information and precision metrology \cite{hansch1975cooling, wineland1978radiation, Dalibard:89, cohen1990new, wineland1998experimental, Chu1998, Cohen-Tannoudji1998, phillips1998nobel,  weiss2017quantum, katori2011optical}. These methods have enabled the exploration of ultracold gases, leading to the realization of Bose-Einstein condensation \cite{anderson1995observation, Davis1995}, where quantum effects manifest on a macroscopic scale, opening new avenues in many-body physics and quantum simulations \cite{bloch2012quantum}. 

In these systems, laser fields replace the conventional thermal heat bath, leading to behaviors that deviate significantly from the predictions of standard statistical physics \cite{Cohen-Tannoudji1998, Bardou2002, Afek2023}. These deviations persist while the laser remains active, maintaining the system out of equilibrium.  Subrecoil laser cooling enables nanokelvin temperatures by exploiting a velocity-dependent fluorescence rate that scales as  $R(v) \propto |v|^\alpha$  for small velocities  $v$ \cite{Bardou1994,esslinger1996subrecoil,Saubamea1999}. This  cooling mechanism allows for the spreading of a packet of atoms, which can be directly measured in experiments. The shape of this packet is strongly influenced by the cooling process. Importantly, the velocity-dependent rate suppresses laser-atom interactions in the low-velocity limit  ($v \to 0$),  effectively slowing atomic motion.  Efficient cooling is achieved when $\alpha > 1$, as the mean lifetime of atoms in the ``dark state"—where at low velocities they effectively cease interacting with the laser—diverges \cite{Bardou2002}. This divergence fundamentally transforms the system's dynamics, disrupting classical concepts of equilibrium and diffusion. Consequently, this unique mechanism facilitates the accumulation of ultracold atoms with non-thermal velocity distributions \cite{Bardou2002,Barkai2021, *barkai2022gas}, as observed in experiments \cite{Aspect1988, Kasevich1992, Davidson1994, Bardou1994, Reichel1995,Marksteiner1996, curtis2001quenched, Prudnikov:03}.

A central challenge in subrecoil-laser cooling is understanding atomic diffusion, which governs both cooling efficiency and positional stability. The fundamental question remains: What is the nature of this diffusion—normal, subdiffusive, or superdiffusive?
Here, we investigate the diffusion dynamics in subrecoil-laser-cooled systems using fundamental stochastic models of subrecoil-laser cooling \cite{Bardou1994, Bardou2002}. In these models,  the time interval between successive velocity changes, known as the flight time, is a key stochastic variable. At low speeds, these intervals are exceptionally long and follow a fat-tailed distribution. Such broad-tailed flight times, within the well-known L\'evy walk approach, are a characteristic typically associated with superdiffusion \cite{Geisel1985, Buldyrev1993, ramos2004levy, klafter2005anomalous, barthelemy2008levy, dieterich2008anomalous, Zaburdaev2015, song2018neuronal, cherstvy2018non}. However, as we show below, under certain conditions, the non-trivial coupling between these prolonged flight times and flight lengths results in subdiffusion instead. \cite{Klafter1987, Shlesinger1987, Alberts2018, *Albers2022, Akimoto2013a, Akimoto2014, zhu2023asymmetric}. 


Since the parameter $\alpha$ is tunable through pulse shaping in experiments \cite{Reichel1995,Bardou2002}, a key question is: What value minimizes diffusion? This is crucial for stabilizing the spreading atomic packet. As we show below, a non-trivial value, $\alpha = 3/2$, leads to the slowest diffusion, maximizing atomic stability. These findings highlight the role of $\alpha$ in controlling transport dynamics. Moreover, they offer a fundamental understanding of anomalous diffusion in subrecoil-laser-cooled systems, laying the groundwork for future advancements in atomic control.

This naturally raises the question: how general is this phenomenon? Furthermore, following Meyer et al.~\cite{Meyer2017,meyer2018anomalous}, we point out at the end of this Letter a novel connection between laser cooling and the Pomeau-Manneville map \cite{PM1979,Manneville1980}---a deterministic model originally introduced to describe intermittency in turbulence, and now central to the study of anomalous diffusion, weak chaos, and aging \cite{Geisel1984, Geisel1985, Gaspard1988,gaspard2005chaos, Barkai2003, Korabel2009,Akimoto2010,Akimoto2012,Akimoto2013}. Thus, our work bridges two disciplines, suggesting that the concept of minimal spreading of concentration profiles may reflect a universal principle in physics.

{\em Models.}---To capture the essential physics of subrecoil laser cooling, we first adopt the stochastic framework  and consider two well-established models introduced by Bardou et al.: the rate-induced L\'evy walk (RILW) and the inhomogeneous random walk (IRW) \cite{Bardou2002}.  The RILW model assumes that velocities are independently randomized, neglecting correlations between successive emissions, while the IRW model retains  these correlations  \cite{Bardou2002}. In subrecoil laser cooling, atomic velocity evolves through stochastic photon emission and absorption,  effectively modeled as a one-dimensional random-walk process with a velocity-dependent jump rate $R(v)$ \cite{Bardou2002, bertin2008laser}.

In the IRW model, velocity updates occur as $v_1 = v_0 + \Delta v$, where  $v_0$  is the initial velocity,   $v_1$  is the post-collision velocity, and  $\Delta v$  is drawn from a Gaussian distribution with mean zero and variance  $\sigma^2$. This model captures the correlation between velocity steps, introducing memory effects in atomic motion. To stabilize the cooling dynamics, we impose reflecting boundaries at \(|v| = v_{\text{max}}\), confining the velocity within the physically meaningful range \([-v_{\text{max}}, v_{\text{max}}]\). This condition reflects experimental setups, which constrain the maximal speed of the atomic motion \cite{gaunt2013bose, esslinger1996subrecoil}. The RILW model, on the other hand, adopts a coarse-grained approach: since multiple spontaneous emissions decorrelate atomic velocity, post-collision velocities are drawn from independent and identically distributed (IID) random variables following a stationary symmetric distribution  $f(v)$. 

In both models, collisions occur at random intervals, where flight times, conditioned on the velocity  $v$, follow an exponential distribution. Later, we will show that when this velocity dependence is removed, the flight-time distribution exhibits a power-law tail.  The fluorescence rate  $R(v)$ \cite{Bardou2002}, which determines the inverse of the mean flight time, is given by
\begin{equation}
	R(v)= \frac{1}{\tau_0} \left| \frac{v}{v_r} \right|^\alpha .
	\label{eq: rate}
\end{equation}
Here, $\alpha$ controls the velocity dependence of the jump rate, and $v_r$ is a characteristic velocity scale for the cooling process. As  $v \to 0$, interactions between atoms and the laser field cease, leading to vanishing jumps. Experimental setups, such as velocity-selective coherent population trapping, typically use $\alpha = 2$ or $4$ to restrict atomic speeds to small values \cite{aspect1989laser, Reichel1995}. 
The parameter $\alpha$ determines how the atomic excitation rate---and hence the fluorescence rate---vanishes at low velocities. In subrecoil laser cooling experiments, this rate is controlled by the spectral shape of the Raman cooling pulses. Specifically, a square pulse produces $\alpha = 2$ \cite{Reichel1995,Bardou2002}, while smoother pulses such as Blackman shapes suppress low-velocity excitation more strongly and result in larger values like $\alpha \approx 4$  \cite{Kasevich1992,Reichel1995}. More generally, other values of $\alpha$ can be realized by tailoring the spectral density near the zero of the pulse power spectrum \cite{Bardou2002}; see Supplemental Material (SM) for additional details \cite{SM}.
The parameter $\tau_0$ in Eq.~\eqref{eq: rate} corresponds to the characteristic photon emission timescale, setting the rate of interaction with the cooling laser \cite{Saubamea1999,Bardou2002}. For simplicity, we set $v_r=1$ and $\tau_0 =1$ in our analysis.

\begin{figure}
	\includegraphics[width=.9\linewidth, angle=0]{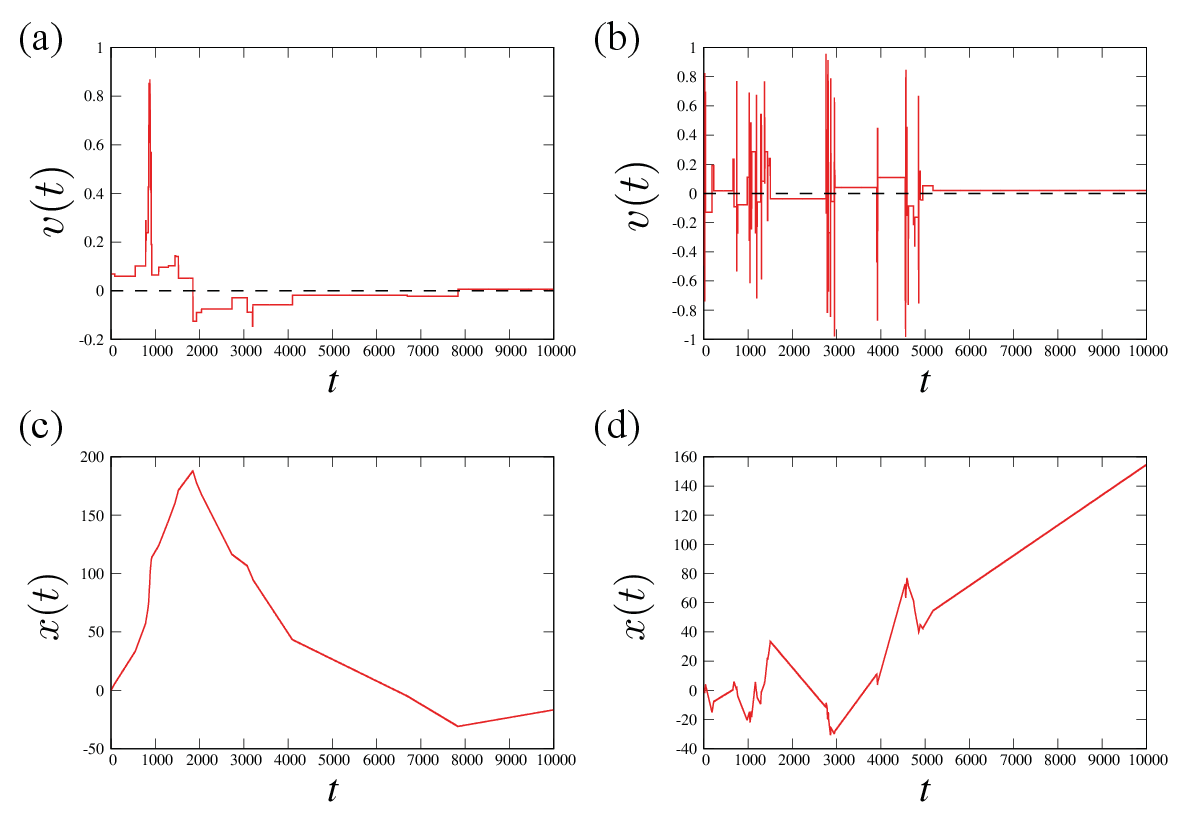}
	\caption {Trajectories of velocity and position in the IRW and RILW models for  $\alpha = 2$. 
	(a), (b) Velocity trajectories in the IRW and RILW models, respectively.
(c), (d) Corresponding position trajectories for (a) and (b), respectively.  
}
	\label{fig:models}
\end{figure}

During a flight, an atom moves at a constant velocity for a random flight time, after which its velocity changes, resembling the dynamics of a L\'evy walk (see Fig.~\ref{fig:models}). 
However, unlike a L\'evy walk, the probability density function (PDF) of flight times depends on velocity $v$ and follows an exponential distribution with a mean flight duration of $1/R(v)$. 
However, since  $v$  evolves stochastically throughout the process, the overall flight time distribution exhibits power-law-like behavior; see Eq.~(\ref{eq: PDF flight time}) below. Physically, these long flights occur predominantly at low speeds, where jump rates are small. This velocity-dependent jump rate plays a crucial role in determining the transport regime, including possible anomalous diffusion. 
Furthermore, in the IRW model, correlated velocity changes between collisions violate the renewal assumption of L\'evy walks, introducing additional complexity into the diffusion process. 

For analytical tractability, we first focus on the RILW model \cite{Bardou2002}, where velocity changes are uncorrelated.
The master equation governing the velocity distribution $\rho(v, t)$ in the RILW model is given by 
\begin{equation}
	\frac{\partial \rho (v,t)}{\partial t} = - R(v) \rho (v,t) +  \int_{-1}^1 R(v') \rho (v',t) f(v) dv',
	\label{eq: master}
\end{equation}
where $v$ represents the velocity immediately after a jump, and $R(v)$ and $f(v)$ are as defined earlier. This equation describes the evolution of the velocity distribution, where the first term accounts for velocity changes due to jumps, and the second term represents the redistribution of velocities according to  $f(v)$, reflecting the stochastic nature of the process.

At long times, the velocity distribution evolves as
\begin{equation}
	\lim_{t \to \infty} Z(t) \rho(v,t) = f(v) |v|^{-\alpha} , 
	\label{master}
\end{equation}
where  $\rho(v,t)$ is the normalized density, and $Z(t)$ is a time-dependent normalization factor \cite{Barkai2021,Bardou1994}. For $\alpha < 1$ and $f(0)\ne0$, $Z(t)$ becomes constant as $t \to \infty$, indicating a steady-state solution. In contrast,  for $1 \leq \alpha$, $Z(t)$ grows as $Z(t)\propto t^{1-\alpha^{-1}}$, signifying the absence of a conventional steady state.  As we will show below,  $\alpha = 1$  also marks a transition in the spatial spreading of particles.

In the RILW model, the position of an atom starting at  $x = 0$  evolves through a series of stochastic flights. Each flight duration  $\tau$  is exponentially distributed with a mean  $1/R(v)$.  During each flight, the atom moves at a constant velocity, resulting in a displacement  $\Delta x = v \tau$. At the end of the flight, the velocity is updated to a new value sampled from the distribution $ f(v)$, and the process repeats.  To illustrate our key results, we set $f(v)=1/2$ for $-1<v<1$ otherwise $f(v)=0$.
Using Eq.~(\ref{eq: rate}), the joint PDF $\psi (\Delta x, \tau)$ of displacement  $\Delta  x$ and flight time $\tau$ is given by 
\begin{equation}
\psi (\Delta  x, \tau) = \int_{-1}^1 \frac{dv}{2} |v|^\alpha e^{-|v|^\alpha \tau}  \delta (\Delta  x-v \tau) .
\label{joint PDF xt}
\end{equation}
The flight-time distribution $\phi(\tau)$ is given by integrating Eq.~\eqref{joint PDF xt} over $\Delta x$:
\begin{equation}
	\phi (\tau) = \frac{\gamma(\frac{1}{\alpha} +1, \tau)}{\alpha} \tau^{-1- \frac{1}{\alpha}}, 
	\label{eq: PDF flight time}
\end{equation}
where $\gamma(\cdot, \tau)$ denotes the incomplete gamma function.  For large  time $\tau$, the flight-time distribution asymptotically behaves as
$\phi (\tau) \sim  \frac{1}{\alpha}\Gamma \left(\tfrac{1}{\alpha}+1 \right) \tau^{-1- \frac{1}{\alpha}}$. Thus, the mean flight time remains finite for  $\alpha < 1$  but diverges for  $\alpha > 1$. We also use the Laplace transform, given by
$\hat{\phi}(s) = 1 - a s^{\frac{1}{\alpha}} + o(s^{\frac{1}{\alpha}})$ for $\alpha>1$, 
where $a=\pi \csc\left(\pi/\alpha\right)/\alpha$. This small $s$ expansion of $\hat{\psi}(s)$ is non-analytical, which reflects the fact that the mean flight time diverges.

The flight-displacement distribution $\varphi(\Delta  x)$ is obtained by the marginal distribution of  Eq.~\eqref{joint PDF xt}.  For $\alpha>1$, this yields
\begin{equation}
\varphi (\Delta  x) = \frac{|\Delta  x|^{-\frac{\alpha}{\alpha -1}}}{2 (\alpha-1)}  \gamma \left(\tfrac{\alpha}{\alpha-1} ,|\Delta  x| \right) .
\label{flight displacement}
\end{equation}
For large displacements, the asymptotic behavior is
\begin{equation}
\varphi (\Delta  x) \sim
\frac{ \Gamma(\frac{\alpha}{\alpha -1} )}{2(\alpha -1)} |\Delta  x|^{-1- \frac{1}{\alpha-1}}\quad (|\Delta  x|\to\infty).
\end{equation}
Note that Eq.~\eqref{flight displacement} is valid only for \( \alpha > 1 \). For \( \alpha < 1 \), the displacement distribution decays exponentially with a power-law prefactor due to the inverse coupling between flight time and displacement (see SM \cite{SM}).
Thus, the mean flight length  $\langle |\Delta x| \rangle$  diverges for  $\alpha \geq 2$, while the second moment  $\langle \Delta x^2 \rangle$  diverges for  $\alpha \geq 3/2$.
Naively, one might expect that since the variance of the jump length diverges at  $\alpha = 3/2$, the system should exhibit superdiffusion. However, as we show below, this expectation is incorrect. Instead,  $\alpha = 3/2$  marks the regime of slowest dispersion, which arises due to the coupling between jump size and flight time.

{\em Anomalous Diffusion: Normal, Subdiffusion, and Superdiffusion.}---To understand the anomalous diffusion of laser-cooled atoms, we derive the mean square displacement (MSD) $\langle x^2 (t)\rangle$  
using the Montroll-Weiss equation \cite{montroll1965random, Shlesinger1982, Akimoto2020-inf}, which relates the propagator $p(x, t)$ to the jump length and  waiting-time distributions. 
The propagator can be expressed as
\begin{equation}
	p(x,t)= \int_{-\infty}^{\infty} d\Delta x \int_{0}^{t} d\tau Q(x-\Delta x,t-\tau) \Psi (\Delta x,\tau) ,
	\label{propageter}
\end{equation}
where $Q(x,t) dx dt$ represents the probability of a random walker arriving at position $x$ at time $t$ after completing a jump, and 
$\Psi(\Delta x, \tau)$  represents the PDF of the displacement during the last flight,  conditioned on the fact that the flight has lasted at least  $\tau$  but is not yet completed.
Since an atom with velocity  $v$  remains in that state for a duration  $\tau$  with probability  $e^{-|v|^\alpha \tau}$, the corresponding displacement distribution is given by
\begin{equation}  
	\Psi (\Delta x, \tau) = \int_{-1}^1 \frac{dv}{2} e^{-|v|^\alpha \tau}  \delta (\Delta x-v \tau) .
\end{equation}
This function describes the distribution of jump sizes occurring between the last collision event and the measurement time  $t$.
Equation~(\ref{propageter}) is a convolution equation, which can be solved using the Fourier transform ($x \to k$) and the Laplace transform ($t \to s$).
The Fourier-Laplace transform  yields the Montroll-Weiss equation \cite{montroll1965random,metzler00}:
\begin{equation}
\hat{p}(k,s) = \frac{\hat{\Psi}(k,s)}{1-\hat{\psi}(k,s)}.
\label{MW eq}
\end{equation}
This equation explicitly yields the Laplace transform of the moments:
\begin{eqnarray}
	\langle \hat{x}^{2n}(s) \rangle = (-1)^n \left. \frac{\partial^{2n}}{\partial k^{2n}} \hat{p}(k,s) \right|_{k=0}, 
	\label{eq: second derivative}
\end{eqnarray}
where $\langle \hat{x}^{2n}(s) \rangle$ denotes the Laplace transform of $\langle x^{2n}(t) \rangle$ with respect to time $t$.

\begin{figure}
\includegraphics[width=.9\linewidth, angle=0]{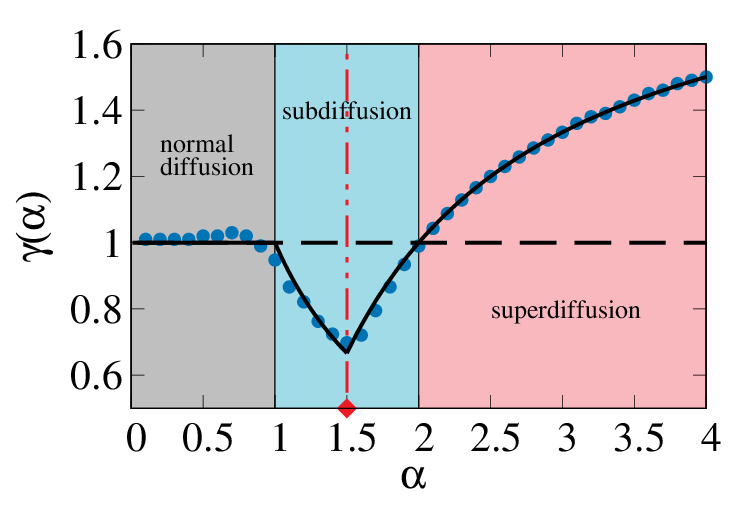}
\caption{Power-law exponent of the MSD as a function of $\alpha$ exhibiting transitions between normal, sub-, and superdiffusion. Symbols are the numerical 
results in the IRW model.  Solid lines represent the theory of the power-law exponent for the RILW model in Eq.~(\ref{eq:coexponent}). To determine the scaling exponent in numerical simulations, we used the final decade of the observation time for fitting, and the slope obtained from this fitting is adopted as the representative value for the analysis.}
\label{fig: power-law exponent MSD}
\end{figure}

In the long-time limit, the leading-order behavior of the MSD is governed by the exponent  $\alpha$ in the jump rate  $R(v)$. 
The MSD reveals distinct scaling regimes dictated by the parameter  $\alpha$. 
Using  Eqs.~(\ref{MW eq}), (\ref{eq: second derivative}) and Tauberian theorems, we obtain the asymptotic behavior of the MSD, which scales as  
$\langle x(t)^2 \rangle \propto t^{\gamma(\alpha)}$
for \( t\to\infty \). The power-law exponent \( \gamma(\alpha) \) is given by
\begin{equation}
	\gamma(\alpha) = \left\{
	\begin{array}{ll}
		1 \quad \left(\alpha<1\right)\\
		\\
		\frac{1}{\alpha} \quad \left(1<\alpha<\frac{3}{2}\right)\\
		\\
		2 - \frac{2}{\alpha} \quad \left(\alpha>\frac{3}{2}\right) .
	\end{array}
	\right.
	\label{eq:coexponent}
\end{equation}
For example, experiments with $\alpha=2$ \cite{Aspect1988,Kasevich1992,Reichel1995} yield $\langle x^2(t)\rangle \sim t$, whereas for $\alpha=4$ \cite{Kasevich1992,Reichel1995} one finds $\langle x^2(t)\rangle \sim t^{3/2}$, indicating a markedly faster spreading in the latter case.
The exponents in Eq.~(\ref{eq:coexponent}) differ from those in the classical L\'evy walk model \cite{Zaburdaev2015}, where only superdiffusion occurs, whereas our model exhibits a broader range of diffusion behaviors, including subdiffusion.
For $\alpha<1$, the MSD increases linearly with time, indicating normal diffusion.  
For $1 < \alpha < 2$, subdiffusion is observed. In particular, for $1 < \alpha < 3/2$, it is characterized by a sublinear growth of the MSD, i.e., $\langle x^2(t) \rangle \propto t^{1/\alpha}$. More precisely, in this regime, the asymptotic behavior of the MSD becomes
\begin{eqnarray}
\langle x^2(t) \rangle \sim  \frac{\langle \Delta x^2 \rangle }{a  \Gamma(1+\frac{1}{\alpha})} t^{\frac{1}{\alpha}} ,
\label{eq: msd1}
\end{eqnarray}
where $\langle \Delta x^2 \rangle$ is the second moment of the flight displacement, see Eq.~\eqref{flight displacement}. 
Subdiffusion persists for $\alpha < 2$. At $\alpha = 2$, normal diffusion is recovered, followed by the onset of superdiffusion for $\alpha > 2$.	
The exact asymptotic behavior for  $\alpha > 3/2$ reads 
\begin{eqnarray}
	\langle x^2(t) \rangle &\sim&  
	\frac{2 (\alpha - 3) \sin\left(\frac{\pi}{\alpha}\right) \csc\left(\frac{3 \pi}{\alpha}\right)}{\alpha \Gamma(3 - \frac{2}{\alpha})} t^{2 - \frac{2}{\alpha}}.
	\label{eq: msd2}
\end{eqnarray}
At $\alpha=3/2$, the MSD exhibits a logarithmic correction (see Suppplemental Material). Consequently, the prefactor in Eqs.~(\ref{eq: msd1}) and (\ref{eq: msd2}) diverges as $\alpha\to 3/2$.
However, the slowest diffusion pattern is found for $\alpha = 3/2$ because the subdiffusive exponent is minimized at this point (see Fig.~\ref{fig: power-law exponent MSD}). 
For $\alpha > 2$, the system enters a superdiffusive regime, where the MSD scales as $\langle x^2(t) \rangle \propto t^{2 - \frac{2}{\alpha}}.$
As we detail later, these behaviors closely mirror those observed in diffusion driven by Pomeau-Manneville maps~\cite{meyer2018anomalous}, underscoring the robustness and generality of our findings.

A basic question is whether similar behavior is found  for the IRW model, where the collisions process exhibits correlations. 
We perform numerical simulations of the IRW model and compare the results with the theory of the RILW model. The simulation starts by assigning an initial velocity \(v_0\) uniformly from \([-1, 1]\). An initial flight time \(\tau\) is then generated from an exponential distribution with mean \(1/|v_0|^\alpha\). In numerical simulations, the observation time is set to \(t = 10^8\), and the number of particles in the sample is \(n = 10^6\).

In long-time asymptotic behaviors, the MSD exhibits a power-law increase, characteristic of anomalous diffusion. Our results demonstrate excellent agreement between the numerical simulations of the IRW model and the theoretical predictions of the RILW model  (see Fig.~\ref{fig: power-law exponent MSD}). Although not shown here, we confirm that numerical simulations of the RILW models with $f(v)=1/2$ for $-1<v<1$ and a Gaussian model $f(v)=\mathcal{N}(0, 1)$ matches with the results of the IRW model,  reinforcing the robustness of our findings.
As shown in Fig.~\ref{fig: power-law exponent MSD}, as  $\alpha$  increases beyond 1, the system transitions from normal diffusion to subdiffusion, underscoring the versatility of the model in capturing a wide range of diffusion behaviors. At  $\alpha = 3/2$, an intricate balance between velocity suppression and prolonged low-velocity states minimizes spatial spreading,  optimizing atomic confinement. This differs fundamentally from standard L\'evy walks, where power-law-distributed flights typically lead to superdiffusion.  
Another transition from subdiffusion ($\gamma(\alpha) < 1$) to superdiffusion ($\gamma(\alpha) > 1$) occurs at  $\alpha = 2$. Despite the mean speed approaching zero at long times,  the system exhibits superdiffusion for  $\alpha > 2$, reflecting the interplay between velocity suppression and long flight times.


{\em Non-Gaussian Parameter and Deviations from Gaussian Behavior.}---It is important to study deviations from Gaussian behavior. For this purpose, we use the non-Gaussian parameter (NGP) \cite{rahman1964correlations}, defined as

\begin{equation}
	{\rm NGP} \equiv \frac{\langle x(t)^4 \rangle}{3 \langle x(t)^2 \rangle^2} -1.
\end{equation}
The convergence of the NGP for $t\to\infty$ implies that the fourth moment of the displacement is proportional to the square of the MSD.
In purely Gaussian processes, the NGP equals zero, as all higher moments are determined solely by the second moment. A positive NGP signals a leptokurtic distribution, exhibiting a sharp central peak and heavier tails, meaning most atoms remain confined while occasional large jumps occur. Conversely, a negative NGP corresponds to a platykurtic distribution, characterized by a broader, flatter profile with more uniform spreading \cite{marchenko2025approach}.
To compute the NGP, we calculate the fourth moment of the displacement using the Montroll-Weiss equation Eq.~(\ref{MW eq}):
\begin{eqnarray}
\langle \hat{x}^4(s) \rangle &=& \left. \frac{\partial^4}{\partial k^4} \hat{p}(k,s) \right|_{k=0},
\label{eq: 4th derivative}
\end{eqnarray}
where $\langle \hat{x}^4(s) \rangle$ denotes the Laplace transform of $\langle x^4(t) \rangle$ with respect to time $t$. We then invert this expression to obtain the time-domain solution. The detailed derivation of the NGP for different regimes of $\alpha$ is provided in the SM \cite{SM}.

\begin{figure}
\includegraphics[width=.9\linewidth, angle=0]{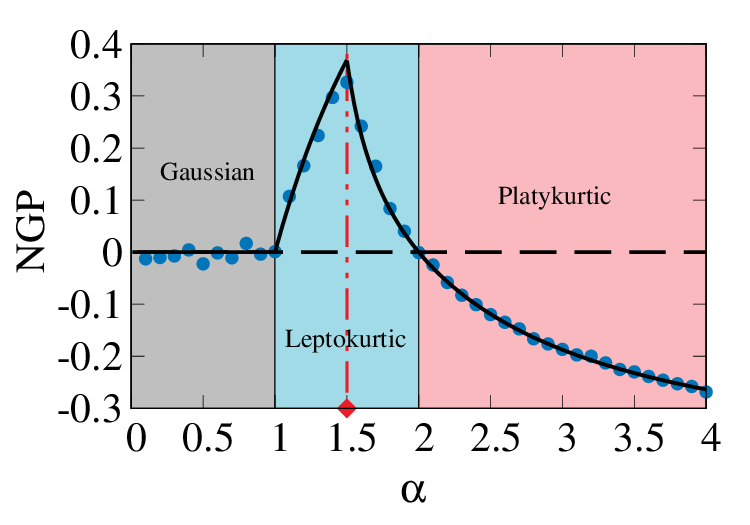}
\caption{Non-Gaussian parameter as a function of $\alpha$. Symbols are the numerical results in the IRW model. Solid lines represent the theory of the NGP for the RILW model. To determine the non-Gaussian parameter in numerical simulations, the final decade of the observation time is used. 
}
\label{fig: NGP}
\end{figure}

The behavior of the NGP across different  $\alpha$-regimes reveals the transition between Gaussian-like and non-Gaussian distributions (see Fig.~\ref{fig: NGP}).
For $\alpha < 1$, the NGP approaches zero in the long-time limit, indicating a Gaussian propagator (see SM for details \cite{SM}). As $\alpha$ increases beyond 1, deviations from Gaussian behavior become pronounced due to diverging higher-order moments of the flight length. In the range $1 < \alpha < 2$, the NGP takes positive values, reflecting a highly peaked and more concentrated atomic packet. As shown in Fig.~\ref{fig: NGP}, at the critical point $\alpha = 3/2$, the NGP attains its peak value, indicating the most sharply peaked and concentrated propagator. This peaked propagator significantly suppresses diffusion, highlighting its relevance for atomic trapping applications. For $\alpha > 3/2$, the NGP decreases and eventually becomes negative, corresponding to a flatter distribution characteristic of ballistic spreading. 

{\em Connection to other models.}—Interestingly, following Ref.~\cite{Meyer2017,meyer2018anomalous}, we uncover a connection between laser cooling and a paradigmatic model in nonlinear dynamics---the Pomeau-Manneville map \cite{PM1979,Manneville1980, Geisel1984, Geisel1985}. This map models systems with intermittent behavior due to marginally unstable fixed points. A control parameter, denoted $z$ (see SM for the detail \cite{SM}), characterizes the nonlinearity near these points.  In studying anomalous spreading driven by the Pomeau-Manneville map, the diffusion is structurally similar to that of our RILW model, with $z$ replaced by $\alpha + 1$. 
Meyer et al. \cite{Meyer2017,meyer2018anomalous} identified a minimal value of the spreading exponent, effectively anticipating our own findings in a completely different physical setting (see SM for a more detailed comparison between the models \cite{SM}).  Note that the dynamics of the two systems are vastly different---one stochastic, the other deterministic---highlighting that the principle underlying the slowdown of diffusion is universal.

Furthermore, our model is structurally related to generalized L\'evy walks~\cite{Alberts2018,*Albers2022} and the annealed transit time model \cite{Massignan2014,Manzo2015,AkimotoYamamoto2016a}, where the flight length or diffusivity is dynamically coupled to the flight or waiting time. 
Such stochastic models typically involve  two independent parameters  to characterize the coupling, in contrast to our framework, which is governed by a single parameter, $\alpha$. 
While similar MSD scaling---potentially featuring a minimum---can emerge in these models, they are generally constructed on ad hoc assumptions and lack a direct microscopic foundation. By contrast, our model derives from a physically motivated mechanism, suggesting that the emergence of a minimal-spreading regime may represent a general and tunable feature of anomalous diffusion.
Recent work has further clarified the relation between L\'evy walks and intermittency in lifted Pomeau-Manneville maps, revealing subtle singular behavior beyond simple L\'evy-walk phenomenology \cite{brevitt2025singularity}.



{\em Conclusion.}---We have demonstrated that anomalous diffusion and the shape of the propagator in subrecoil-laser-cooled atoms are tunable via the velocity-dependent fluorescence rate. By varying the parameter $\alpha$, we identify the transitions between normal diffusion, subdiffusion, and superdiffusion. Our findings establish  $\alpha = 3/2$  as the optimal point where diffusion is minimized due to an intricate interplay between velocity suppression and prolonged low-speed flight times. Notably, the NGP, a measure of deviation from normal diffusion, also exhibits a pronounced peak at $\alpha = 3/2$ (see Fig.~\ref{fig: NGP}), coinciding with the minimum in spatial spreading. Our results apply to a broad class of subrecoil cooling schemes in which the excitation rate can be tuned via pulse shaping. 
Our approach opens avenues for precision control of laser-cooled atoms, with potential applications in atomic trapping and quantum technologies.

{\em Data Availability Statement.}---The data supporting the findings of this study are available from the corresponding author upon reasonable request \cite{data_availability}.

We would like to express our deep gratitude to Prof. G\"unter Radons for his invaluable contributions to this work. We are deeply saddened by his recent passing. 
We thank Daiki Kanbe for valuable contributions to the construction of the $\alpha$ pulse.
T.A. was supported by JSPS Grant-in-Aid for Scientific Research (No.~C 21K033920). The support of Israel Science Foundation's grant 1614/21 is acknowledged (EB).


%

\end{document}